\documentclass[preprint,12pt]{elsarticle}
\usepackage{amsmath}
\usepackage{subcaption}
\newcommand{\angstrom}{\text{\normalfont\AA}}

\usepackage{amssymb}
\journal{Planetary and Space Science}

\begin{document}

\begin{frontmatter}



\title{Formation and properties of astrophysical carbonaceous dust}


\author[labmauney]{Christopher M. Mauney\fnref{osuref}}
\author[lablazzati]{Davide Lazzati\fnref{osuref}}
\address[labmauney]{mauneyc@oregonstate.edu}
\address[lablazzati]{lazzatid@science.oregonstate.edu}
\fntext[osuref]{Oregon State University}

\begin{abstract}
The classical theory of grain nucleation suffers from both theoretical and predictive deficiencies.  We strive to alleviate these deficiencies in our understanding of dust formation and growth by utilizing an atomistic model of nucleation. Carbon cluster geometries are determined with a  set of global minimization algorithms.  Using density functional theory, the binding energies of carbon clusters from $n=2$ to $n=99$ are then calculated.  These energies are used to calculate the critical size and nucleation rate of carbon clusters.

We find that the critical cluster size is largely determined by the changes in geometry of the clusters. Clusters with size $n=27$ and $n=8$, roughly corresponding to the transition from ring-to-fullerene geometry and chain-to-ring geometry respectively, are the critical sizes across the range of temperature and saturation where nucleation is significant.  In contrast to the classical theory, nucleation is enhanced at low-temperatures, and suppressed at high temperatures. These results will be applied to a modified chemical evolution code using results from supernova simulations.

\end{abstract}


\end{frontmatter}


\section{Introduction}
\label{sec:intro}
Core-collapse supernovae (CCSNe) provide an excellent candidate for the source of astrophysical dust.  In particular, CCSNe offer the best explanation for the presence of dust in galaxies of the young universe, where other dust production environments would not have formed at such early times \citep{Bertoldi03, Wang08}. However, the formation of dust in CCSNe remains a poorly understood event. Using the classical approach to nucleation, a large amount of dust mass is condensed very quickly within a year after explosion.  Alternatively, kinetics models based on chemical networks manifest slower dust growth and less dust production \citep{Cherchneff09}.  These predictions both suffer from inherent drawbacks in the respective theories used.

%Dust mass predictions of supernova explosions using the classical theory of dust nucleation and growth have over-predicted the rate of dust formation by many orders of magnitude \citep{Todini01,Fallest11,Gall14}.
%
The classical theory of nucleation (CNT) is well known to be deficient, especially when applied to astrophysical dust where nucleation occurs at small cluster sizes \citep{Donn85,Lazzati08}.  The capillary approximation, an integral component of CNT, assumes clusters have a well-defined and simple surface used to predict the binding energies of small clusters.  This leads to model that discards the interactions of dust with the surrounding environment.  Furthermore, the capillary approximation does not adequately handle the discrete nature of small clusters, where addition of monomers can result in radically different properties of cluster energies and geometries.  In contrast, a kinetics based approach disregards the energetic stability of critical clusters, and does not capture the exponential growth of dust grains.  This results in dust formation beginning too late and grain growth that proceeds too slow \citep{Cherchneff09, Sarangi13}.

%In opposition, a purely kinetics based approach to grain formation generally lacks a physically real chemical reaction network necessary to correctly model the efficient formation of dust grains.  By disregarding the energetic stability of critical clusters, a kinetics based approach does not capture the exponential growth of dust grains, and as a result begins dust formation too late and grain growth proceeds too slow \citep{Cherchneff09, Sarangi13}.
%

Early observations of SN1987A indicated an amount of dust ranging from 10$^{-4}$ to 10$^{-3}$ M$_{\odot}$ at around 1000 days after explosion \citep{Wooden93}.  Later far infrared and sub-mm studies with the \textit{Herchel Space Observatory} approximately 20 years later indicated increased dust mass on the order of 1 M$_{\odot}$ \citep{Wesson14}.  This apparently large and cold dust mass was analyzed in \citep{Matsuura15} by fitting the emission spectrum of various dust compositions, and found that the likely composition is mainly amorphous carbon and silicates totaling approximately 0.8 M$_{\odot}$.  The growth of dust in the interval was attributed to cold accretion onto seed grains formed at early times.  However, Dwek \& Arendt \citep{Dwek15} have proposed rapid formation of a large amount of dust at earlier times (around 2-3 years), and only a small percentage of the total dust mass in later times due to cold accretion.  This study also confirmed larger amounts of silicate over carbon dust in the ejecta.

Resolving these differing observations in theoretical models remains an elusive goal. The question of when and how dust grows, whether in large amounts shortly after the SN explosion or slow and cold growth during years spent in the remnant phase.  Sarangi \& Cherchneff \cite{Sarangi15} describe a model where small precursors of dust molecules are formed in a chemical kinetics network, and thereafter grow by coagulation.  They found a complicated multi-layered process where dust growth is highly dependent on formation history and morphology of the remnant.  Nucleation of dust grains still constrains the eventual masses of dust, and although allowing accretion onto dust is included in the model it does not appear in their model to account for any significant of dust in the late-stage remnant.

%Recent observations of dust mass and composition in CCNSe, particularly SN 1987A, have shown a large amount of dust formation exceeding measurements of earlier epochs by several orders of magnitude\citep{Matsuura14}.  The low amount of dust in the earlier measurements has been confirmed \citep{Wesson14} and implies the continuing formation and growth of dust matter in the outflow.

The advent of many-body quantum techniques, in particular Density Functional Theory (DFT) have allowed us to re-examine a more fundamental model for nucleation.  Our approach discards the capillary approximation in grain formation, and instead employs an atomistic approach.  We sketch the details of our calculations using DFT to carbon clusters below.  As physically real grain nucleation does not take place \emph{in vacuo}, we propose a model of grain growth where critical clusters nucleate and grow within a standard chemical network.  In this manner, we can modulate the runaway growth of dust grains by allowing monomer material the freedom to transit between the vapor (which supplies the dust material) and small molecules.

\section{Methods}
\label{sec:methods}
The procedure to determine cluster binding energies is as follows: first we find the minimum energy candidates using global minimization techniques; then we use DFT to select the lowest energy candidate from the previous step.  We briefly describe these steps below.

\subsection{Determining cluster geometries}
To properly determine the ground state binding energy of a cluster, the geometry (configuration of the atoms) of the cluster must first be found.  This is not a trivial undertaking, and is an ongoing area of active research.\citep{Bauschlicher10,Goumans12}.  In truth, no method exists to determine the exact ground state configuration of a molecular cluster, and furthermore no method exists that provides an exhaustive search of the potential energy surface (PES).  However, our selection of methods does provide a reasonably accurate approximation to complex geometries of clusters.

We use a combination of simulated annealing, basin hopping \citep{Wales97}, and minima hopping \citep{Goedecker04}.  Our \emph{cost function} in these searches is the empirical potential of Brenner \cite{Brenner90} for hydrocarbons.  For small cluster size ($n<20$), simulated annealing is sufficient to find ground-state candidates in a reasonable amount of time.  For larger sizes ($n>20$) we use a sequence of basin hopping and minima hopping with progressively refined parameter inputs. This procedure naturally gives several possible candidates for the ground state energy.  We then use DFT to relax the candidate clusters into the local energy minima, and extract the lowest energy candidate as the true ground state.

This portion of our work is not intended to provide a definitive collection of the configurations of carbon clusters under study, but rather to provide both an accurate description of carbon cluster energetics at this scale and to build a software framework to facilitate research of other molecular clusters of different species that are less well studied (for instance, astrophysical silicates).  Our previous paper on this subject \citep{Mauney15} goes into a more complete discussion of our results and how they compare to similar findings in other work.

\subsection{Density Fucntional Theory}
The key of our approach to nucleation is the inclusion of DFT to calculate the binding energies of clusters.  We use the open-source software \textit{Quantum Espresso} (QE) \cite{Giannozzi09} to calculate the ground state binding energies and relax the cluster to the local minima state.  QE uses a plane-wave basis to solve the coupled Kohn-Sahm system of equations to determine the ground state number density.

We selected the generalized gradient approximation of Perdew-Burke-Ernzerhof \cite{Perdew96} as our exchange-correlation energy functional.  This functional appears to most accurately describe the behavior of cluster geometry transitions in the critical range $20 \le n \le 30$.  QE uses an effective core potential (pseudopotential) which, rather than modeling the entire atom of nucleus and electrons, only considers the valence electrons and replaces the core electrons and nucleus with a pseudoatom.  This pseudoatom exerts a pseudopotential on the active valence electrons.

QE uses an iterative self-consistent technique for solving the coupled
Kohn-Sahm equations. Self-consistency is achieved when the relative
convergence is less than \(10^{-6}\). Mixing of the iterative number
densities is done using Broyden mixing with a mixing coefficient of
\(\alpha = 0.7\).  An ultrasoft pseudopotential was chosen to minimize the kinetic energy cutoff required for convergence. As QE uses a plane-wave basis, supercell optimization is required to minimize spurious periodic effects.  We find that placing a vacuum of 6 $\angstrom$ on all sides is sufficient to suppress these effects.

Candidate configurations are subjected to a local relaxation using BFGS minimization, as the geometry found in the candidate search will likely be altered by quantum effects (e.g., C$_6$ undergoes distortion due to the Jahn-Teller effect\cite{Saito99}).

\section{Theory}
\label{sec:Theory}
\subsection{Nucleation}
Energies from DFT are supplied to an atomistic formulation of nucleation.  The nucleation rate, defined as the number of clusters nucleated in the saturated gas per unit volume per unit time is given as
\begin{equation} \label{eq:nucl_jrate0}
J = \left[z \gamma \sigma^* C_1^2 \sqrt{\frac{kT}{2 \pi m_0}}\right]  e^{-\Delta G_n^*/kT}
\end{equation}
where $z$ is the Zeldovich factor accounting for back-decay of critical clusters, $\gamma$ is the probability of an incoming monomer to bind to the cluster, $\sigma^*$ is the surface area of the critical cluster, $C_1$ is the concentration of monomers in the vapor, and $m_0$ is the monomer mass.  The key term in this equation is $\Delta G_n^*$, the energy barrier between the vapor phase and the dust phase.  The $*$ designates the value to be taken at the critical size, which is defined as the size where $\Delta G_n$ is maximum.

In the capillarity approximation, $\Delta G_n$ is written as
\begin{equation} \label{eq:nucl_gex_cnt}
\Delta G_{n} = -n \Delta \mu + \sigma c (n v_0)^{\frac{2}{3}}
\end{equation}
The capillarity approximation is expressed in the second term, which gives the excess energy of cluster formation as proportional to the bulk surface tension $\sigma$.

Our approach is to discard this approximation, and to use the exact formation
\begin{equation} \label{eq:nucl_gex}
\Delta G_{n} = -n \Delta \mu + (\lambda n - E_n)
\end{equation}
$E_n$ is the binding energy, which we have determined using DFT. $n\lambda$ is the work necessary to transfer $n$ particles from
the old vapor phase to the bulk of the solid dust phase.

The nucleation rate Eq.(\ref{eq:nucl_jrate0}) is determined by the critical size $n^*$, where $\Delta G_n$ is at a maximum.  In the classical theory, this can be determined analytically, however in the atomistic case we are forced to determine it by inspection of the values.

For a more detailed overview of the theory of nucleation, see references \cite{KashchievTB}.  More on the differences of between the atomistic and classical nucleation can be found in\cite{Mauney15}.

\subsection{Chemical evolution of supernova remnant}
To utilize this work in the prediction the dust mass, we have constructed code to model the chemical evolution of a supernova outflow.  A chemical network is defined as formation or destruction of chemical species

\begin{align}
X + Y \rightarrow Z\\
Z \rightarrow X + Y
\end{align}
Each reaction is associated with a set of fitted parameters $(\alpha, \beta, \gamma)$ that define the reaction rate between species $i,j$ in the form of an Arrhenius expression

\begin{equation}
k_{i,j}= \alpha \left(\frac{T}{300}\right)^{\beta} \exp(-\gamma/T)
\end{equation}

Following the model given in \citep{Cherchneff09}, The code integrates a set of non-linear first-order equations that define the kinetics of the system.

\begin{equation}
\frac{d}{dt}n_i = \sum_j k_{i,j} n_i n_j - \sum_m k_{i,m} n_i n_m
\label{eq:chemkinetic_0}
\end{equation}

To follow the number density of gas particles within dust $n_{i,grain}$, where $i$ represents the species of dust grain, we define the rate of dust transfer per unit volume from the gas-phase to the dust phase as
\begin{align}
\Delta_i &= J_n n^* + n_i (n_{i,grain} v_0)^{2/3} \sqrt{\frac{kT}{2 \pi m_0}}
\end{align}
%
%update%
The number density $n_{i,grain}$ represents the amount of a species material that has condensed into dust.  For example, $n_{C, grain}$ counts the number of carbon atoms that are locked into the growing dust grains.  This quantity is used to track the total dust mass of various species, as well to determine the rate of monomer attachment onto the surface of the grain. $v_0$ is the monomer volume.

The original equations of chemical kinetics Eq.(\ref{eq:chemkinetic_0}) are modified to include loss of material to, and growth of, dust
\begin{align}
&\frac{d}{dt}n_i = \sum_j k_{i,j} n_i n_j - \sum_m k_{i,m} n_i n_m - \Delta_i\\
&\frac{d}{dt}n_{i,grain} = \Delta_i
\end{align}

We use an RK4 integration scheme with a variable time-step to maintain number conservation.  Our initial state is from a 1D nucleosysthesis simulation of a supernova \citep{Lazzati14}.  Values of temperature and saturation are interpolated from the data using a cubic spline.  Nucleation rates and critical sizes are also found from interpolation of calculations done beforehand in order to save the significant computational cost of calculating these values during the integration steps.  Our chemical network is a modified list of reactions from \cite{Cherchneff09}, removing large pure carbon molecules under the assumption that the formation of these molecules are the precursor(or post-cursor) to critical clusters and are already captured in our model of grain growth.

It may be remarked that an ideal chemical network that captures sufficient grain growth up to and beyond the critical size (e.g., successive applications of the reaction C$_n$ + C $\rightarrow$ C$_{n+1}$) would be preferable to nucleating critical clusters directly from the gas phase.  However, the practical difficulties in constructing such a network that would be applicable to interesting environments are significant.  The reaction parameters will need to be evaluated across several temperature regimes, and many heretofore unknown or poorly understood reactions will require detailed investigation. In the absence of a complete network, nonphysical results from kinetics calculations are unavoidable.  Furthermore, as the reaction rates will need to be calculated at each time-step of the integration, and the abundances of many intermediary species will need to be stored, a massive reaction network will be prohibitively taxing on computational resources.

\section{Results}
\begin{figure}
	\centering
	\begin{subfigure}[b]{0.45\textwidth}
		\includegraphics[width=\textwidth]{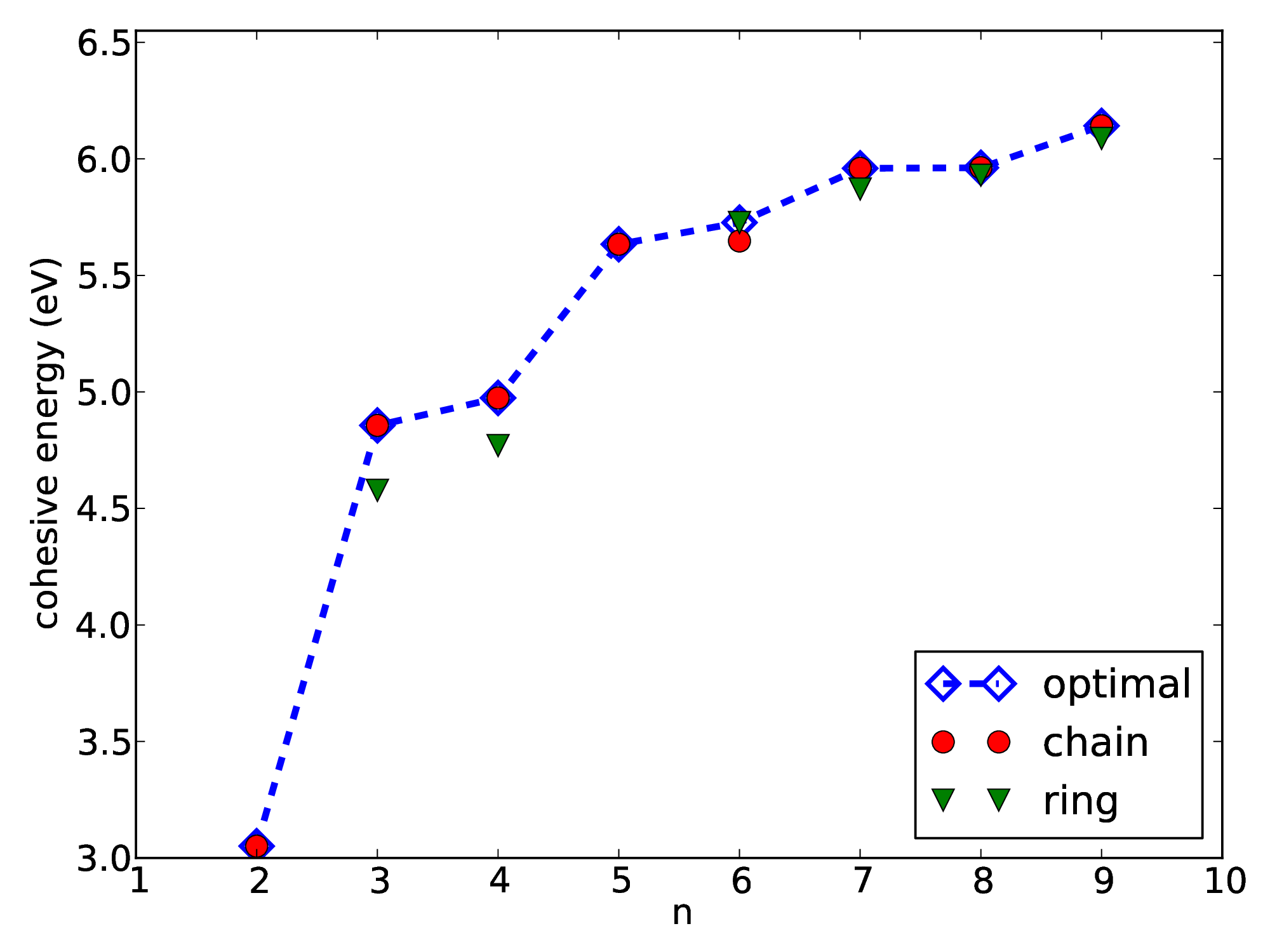}
		\caption{C$_2$ to C$_9$}
		\label{fig:c2c9}
	\end{subfigure}
	\begin{subfigure}[b]{0.45\textwidth}
		\includegraphics[width=\textwidth]{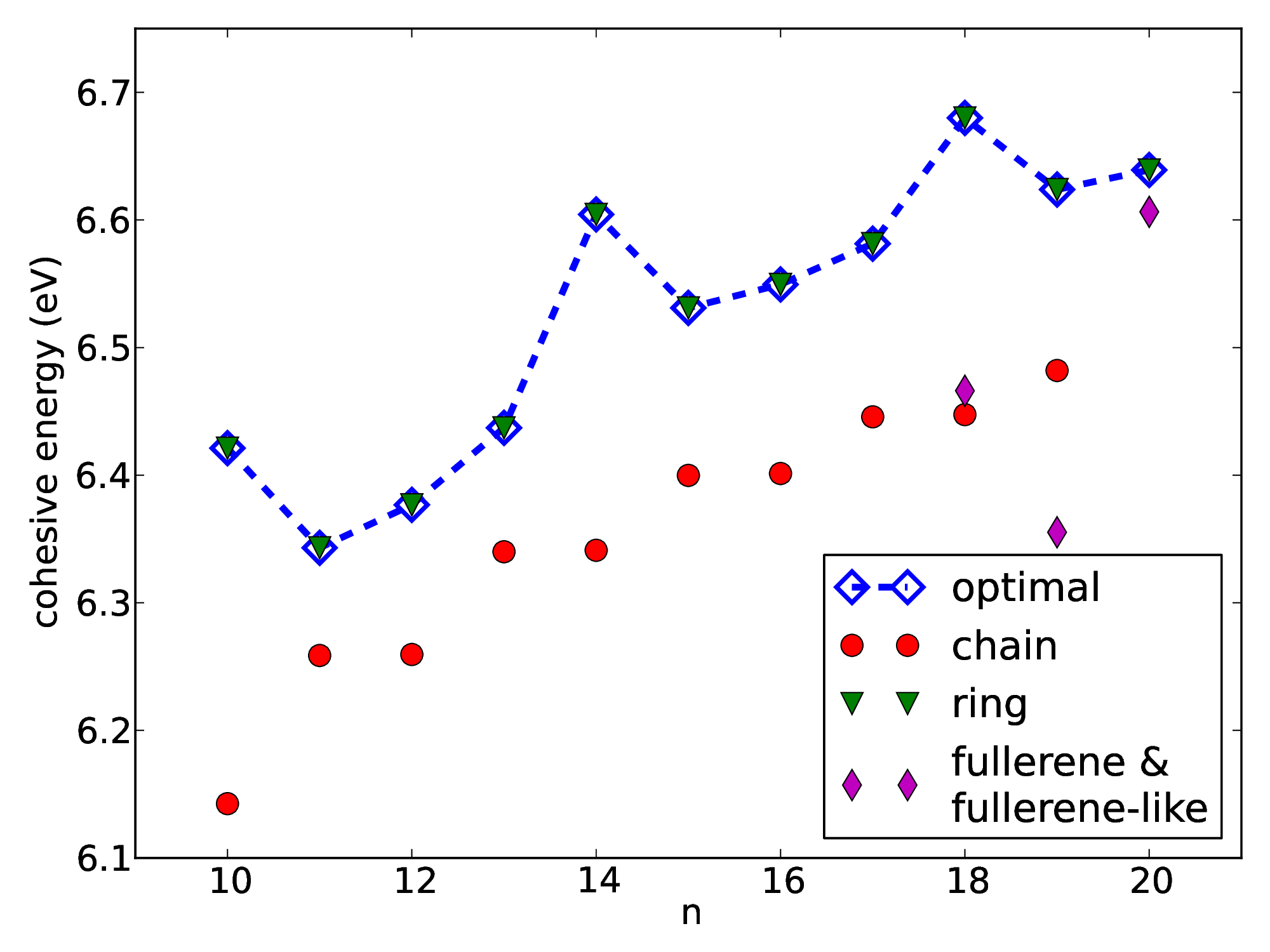}
			\caption{C$_{10}$ to C$_{20}$}
		\label{fig:c10c20}
	\end{subfigure}
	\begin{subfigure}[b]{0.45\textwidth}
		\includegraphics[width=\textwidth]{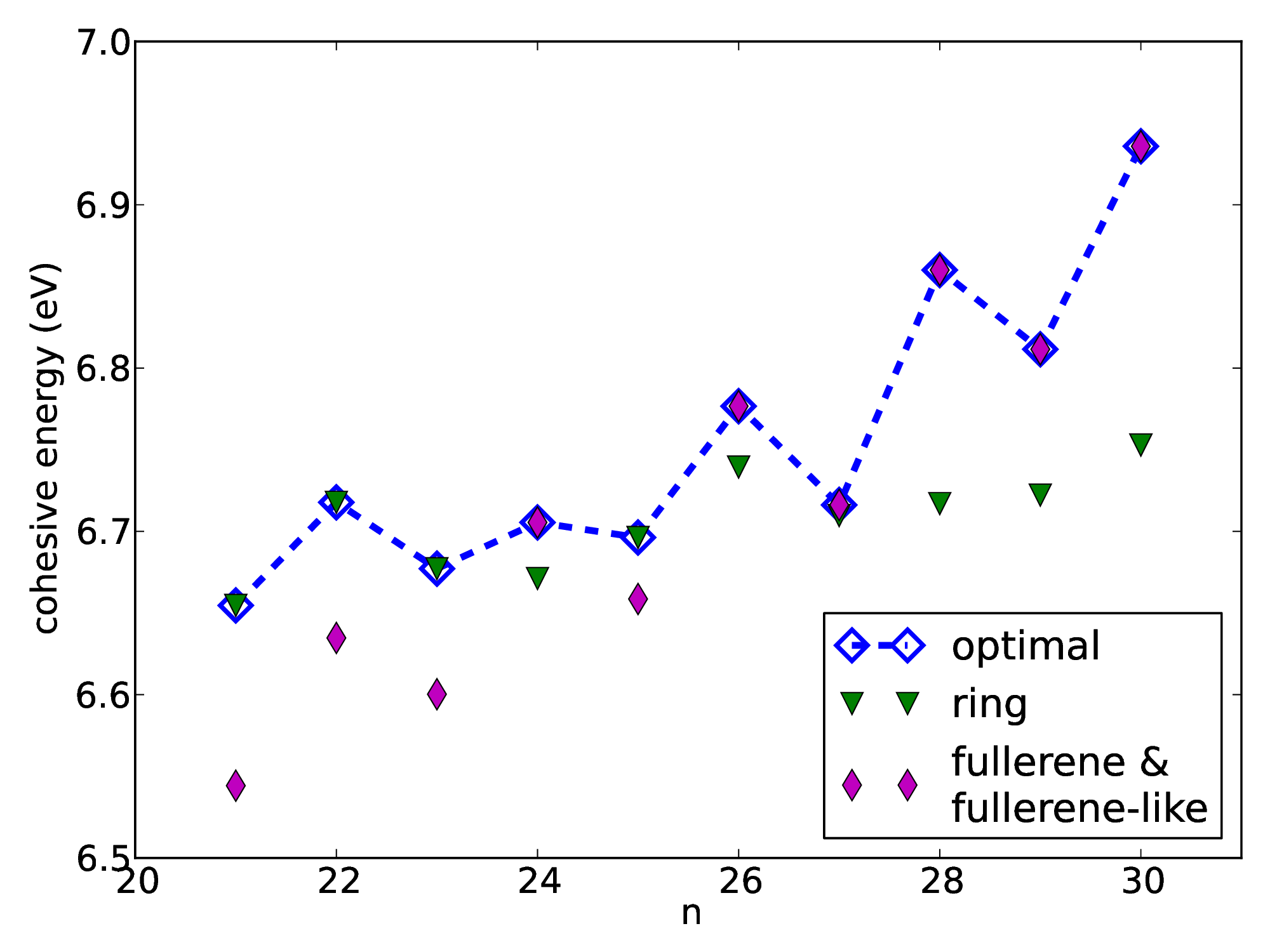}
		\caption{C$_{21}$ to C$_{30}$}
		\label{fig:c21c30}
	\end{subfigure}
	\begin{subfigure}[b]{0.45\textwidth}
		\includegraphics[width=\textwidth]{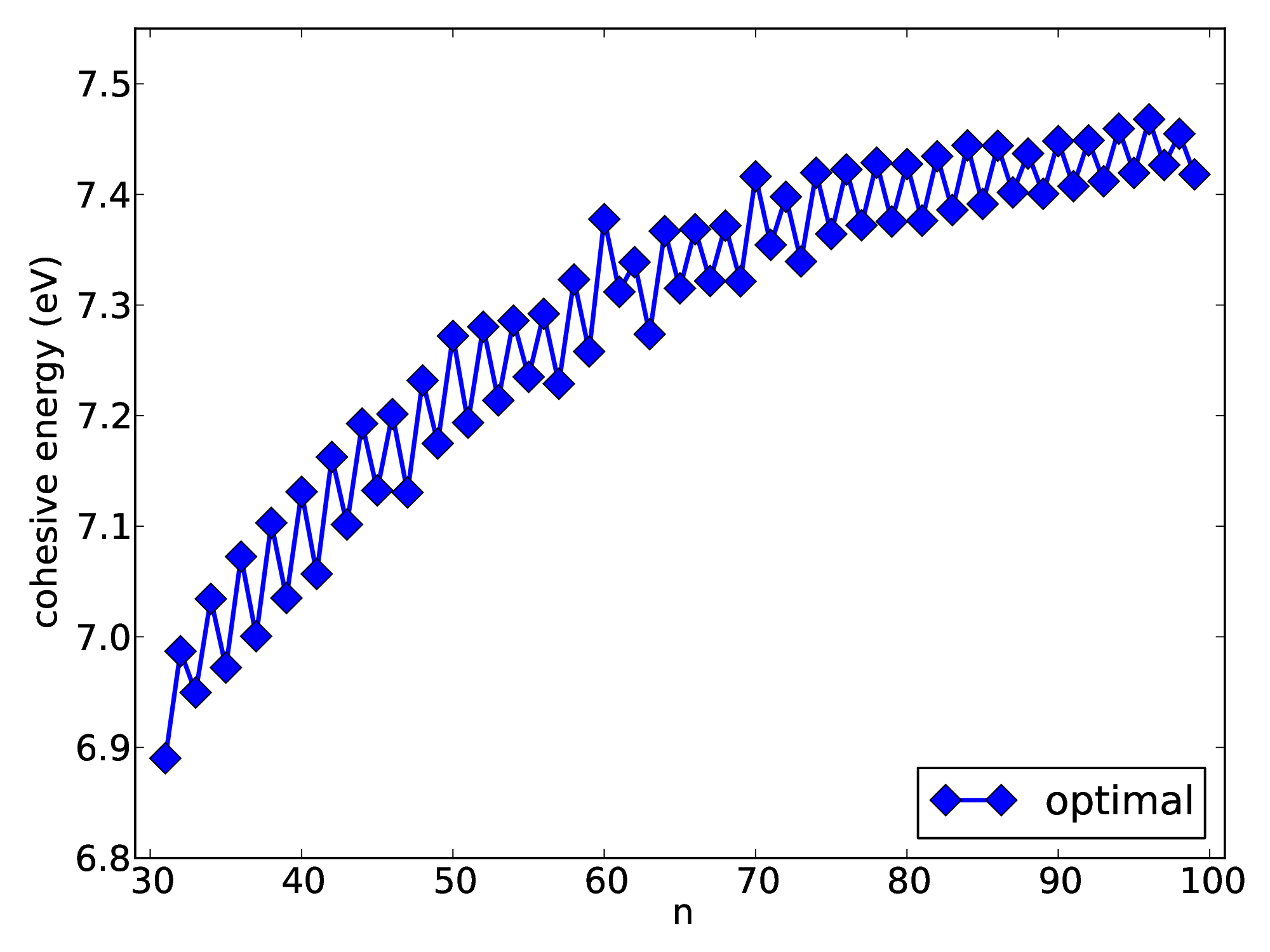}
		\caption{C$_{31}$ to C$_{99}$}
		\label{fig:c31c99}
	\end{subfigure}
	\caption{Cohesive energies ($E_b / n$) of carbon clusters.  Different candidate configurations (chain, ring, fullerene) are plotted along with the optimal geometry.}
	\label{fig:c2c99}
\end{figure}

The binding energies of carbon clusters from DFT calculations are plotted in Figure \ref{fig:c2c99}.  Geometry transitions occur at $n=9$ and $n=27$ for the chain-to-ring and ring-to-fullerene, respectively.  These sizes, along with a subtler variation around $n=6$ (where the optimal cluster transitions from chain to ring back to chain) are seen to coincide with the critical size (see below).

\begin{figure}
	\includegraphics[width=0.5\textwidth]{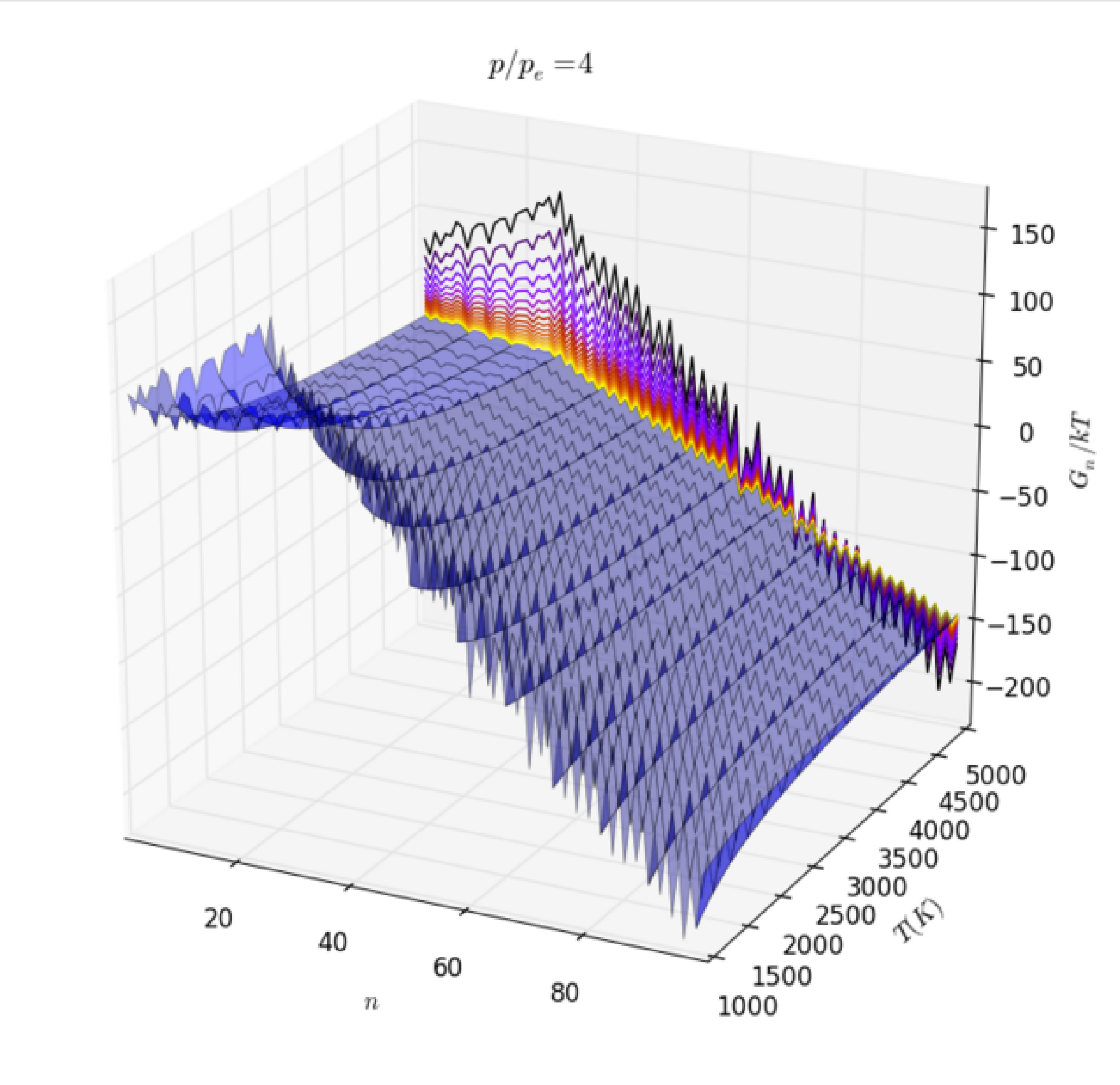}
	\caption{The work of cluster formation $\Delta G_n$ at a sample saturation $S=4.0$, plotted for various temperatures.  The plots for $\Delta G_n$ at various temperatures are projected on the back.  The maximum value of $\Delta G_n$ determines the critical size. The higher temperature, the lower the barrier to cross into the new phase, giving lower critical sizes.}
	\label{fig:wcf}
\end{figure}

An example work of cluster formation $\Delta G_n$ is given in Figure \ref{fig:wcf}.  There is a pronounced peak at the transition of ring-to-fullerene clusters at low temperatures, steadily decreasing as the temperature rises.  The maximum value can be seen to shift towards lower values of $n$, although this changes discretely from $n=27$ to $n=8, 4, 2$.  This is expected to happen at larger temperatures, where the energy barrier between phases is lessened.

\begin{figure}
	\includegraphics[width=0.5\textwidth]{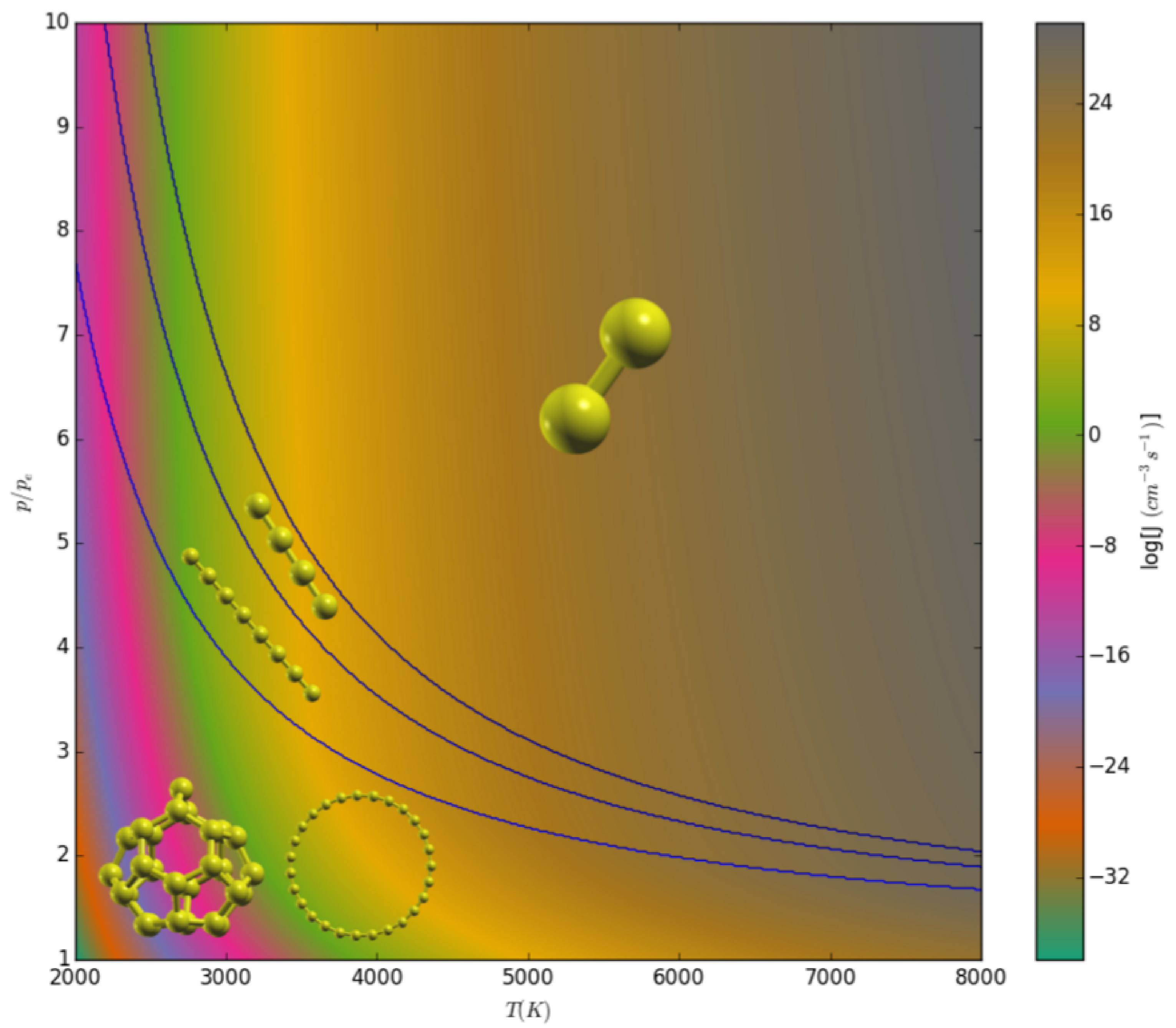}
	\caption{The nucleation rate $J$ for temperature, saturations in the range of signficant nucleation.  The blue lines differentiate the regions with different critical sizes, with critcal clusters are overlayed. The critical size $n=27$ has an ambiguous structure between the ring/fullerene geometry.}
	\label{fig:nuclj}
\end{figure}

The nucleation rates along with critical cluster sizes are are plotted in Figure \ref{fig:nuclj}.  We can see for large values of temperature and saturation the critical size is the dimer C$_2$, indicating that the system is very eager to transition from the gas to solid phase.  Lower values temperature and saturation have a more deliberate path between phases, allowing larger clusters to form before runaway growth of the new phase begins.

\begin{figure}
	\includegraphics[width=0.95\textwidth]{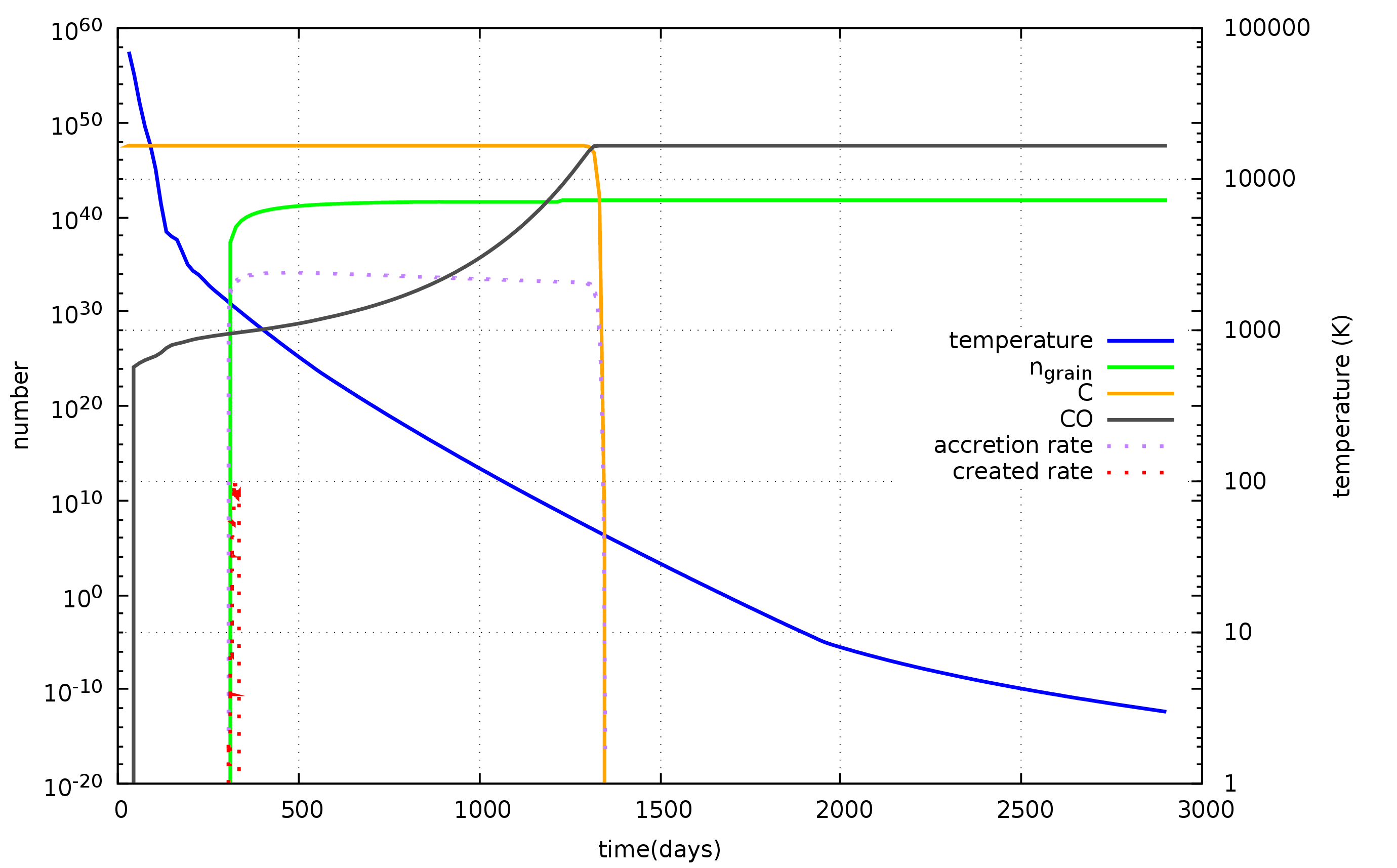}
	\caption{Chemical evolution of a supernova region. The creation/accretion \emph{rate} (per unit time) of grains is plotted alongside the total number of carbon, carbon grains and CO.}
	\label{fig:kinetic}
\end{figure}

An example of dust growth for a carbon-poor region of a supernova is given in Figure \ref{fig:kinetic}, which is located in the first grid cell of the model.  
%update
We use a 15 M$_\odot$ 1-D model for the initial composition of the outflow.  Details on the supernova model we used and the results of applying classical nucleation can be found in \citep{Lazzati14}.
The window of nucleation is small, as seen in the plot of created dust grains, but these begin to quickly accumulate gas monomers.  Initially the majority of the carbon atoms go into the growth of grains, but then saturates.  CO begins to become the preferred repository of carbon and eventually consumes the remaining carbon.  
%update
The interaction of carbon with other elements and molecules is not followed in this case, but can be easily accounted for by including more reactions with varied species in the chemical kinetics network.
%

%update
Development of our model and comparison with astrophysical observations is ongoing.  We are currently applying our approach to carbon - cluster energetics, DFT analysis, and nucleation rate calculation - to silicate clusters.  As the atomic building blocks of these clusters are more varied, consisting of silicon, oxygen, magnesium and iron, potentials in both classical and quantum calculations for these clusters are more varied, and nucleation pathways are much more intricate than in the carbon-only case.  Furthermore, there is a paucity of results in the literature for many of the clusters required for a full description of astrophysical silicate nucleation.  We hope to alleviate this deficiency, and in doing so provide a fuller description of dust formation in our model.

\section{Conclusions}
\label{sec:conclusion}
Beginning with a comprehensive study of carbon cluster geometries using global minimization techniques, and applying density functional calculations to the results, we have determined the binding energies of carbon clusters from a minimum of initial assumptions.  We then applied these binding energies to the atomistic formulation of nucleation theory to produce updated nucleation rates for these carbon clusters. These rates are integrated into a kinetics code to model nucleation in a fully chemical environment, and using data from 1D core-collapse supernova simulations to model grain growth.

The results from our nucleation calculations show that geometry transitions of clusters, notably at $n=27$ and $n=8$ are the preferred critical sizes, with the critical size decreasing with temperature as in the classical case but only falling on or near discrete clusters size about to undergo a transition.  In the classical theory the critical size increases continuously as temperatures fall, leading to a larger barrier to traverse to the new phase.  Thus, our results with the atomistic model show an enhancement of nucleation at lower temperatures.  A similar but opposite case in the regime of high temperature also holds - larger critical sizes in the atomistic case suppresses nucleation rates in relation to the classical theory.

We are continuing to improve our code for forming and growing grains in a full chemical environment. Our progress is towards taking the results of 3D supernova nucleosynthesis simulations that also grow silicate- and iron-bearing grains.  We anticipate to be able to apply our global minimization and DFT procedures to these cases in the future.  In the mean time, we are using previously done \emph{ab-initio} calculations on silicates for our nucleation and dust growth calculations \cite{Goumans12}.

This work was supported in part by NSF grants
1150365-AST and 1461362-AST.





\end{document}